\documentclass[mathleft
]{an}
\usepackage{graphicx}
\usepackage{rotating}
\usepackage{lscape}
\usepackage{longtable}
\usepackage{times}
\usepackage{fge}
\begin{document}

\Pagespan{789}{}
\Yearpublication{2011}%
\Yearsubmission{2010}%
\Month{11}%
\Volume{999}%
\Issue{88}%

\sloppy

\title{Supernova SN 1006 in two historic Yemeni reports}

\author{W. Rada\inst{1}
\and
R. Neuh\"auser\inst{2} \thanks{Corresponding author: \email{rne@astro.uni-jena.de}}
}

\titlerunning{SN 1006 from Yemen}
\authorrunning{Rada \& Neuh\"auser}

\institute{
Hilla University College, Babylon P.O.B. (386), Iraq, (e-mail: wsrada@hotmail.com)
\and
Astrophysikalisches Institut und Universit\"ats-Sternwarte, FSU Jena,
Schillerg\"a\ss chen 2-3, D-07745 Jena, Germany (e-mail: rne@astro.uni-jena.de)
}

\received{2014 Nov 25} 
\accepted{2015 Feb 25}
\publonline{2015 Apr 24}


\abstract{We present two Arabic texts of historic observations of supernova SN 1006 from Yemen
as reported by al-Yam\={a}n{\={\i}} and Ibn al-Dayba$^{c}$ (14th to 16th century AD). 
An English translation of the report by the latter was given before (Stephenson \& Green 2002),
but the original Arabic text was not yet published.
In addition, we present for the first time the earlier report, also from Yemen, namely by al-Yam\={a}n{\={\i}}
in its original Arabic and with our English translation.
It is quite obvious that the report by Ibn al-Dayba$^{c}$ is based on the report by al-Yam\={a}n{\={\i}}
(or a common source), but the earlier report by al-Yam\={a}n{\={\i}} is more detailed and in better (Arabic) language.
We discuss in detail the dating of these observations.
The most striking difference to other reports about SN 1006 is the apparent early discovery in Yemen in the evening of 
{\em 15th of Rajab} of the year 396h (i.e. AD 1006 Apr $17 \pm 2$ on the Julian calendar),
as reported by both al-Yam\={a}n{\={\i}} and Ibn al-Dayba$^{c}$, 
i.e. $\sim 1.5$ weeks earlier than the otherwise earliest known reports.
We also briefly discuss other information from the Yemeni reports 
on brightness, light curve, duration of visibility, location, stationarity, and color.
} 

\maketitle

\section{Introduction: Supernova SN 1006}

Historic observations of supernovae (SN) are useful for 
identification and age dating of SN remnants (SNR) and neutron stars.
There were some ten such SNe observed in the last two millenia,
all from China, three from Europe, and two from Arabia
(see Stephenson \& Green 2002, henceforth SG02).

The previously known Arabic reports about SN 1006 present a lot of detailed information (Goldstein 1965):
From the ecliptic longitude of the SN as given by $^{\rm c}$Al\={\i} ibn Ri\d{d}w\={a}n
(and an error bar from $^{\rm c}$Al\={\i} ibn Ri\d{d}w\={a}n's assumed measurement precision)
together with the declination limit from a St. Gallen observation of this SN (Goldstein 1965) and the Chinese right ascension range
(from Chinese {\em lunar lodge}), it was possible to constrain the location of the SN and to identify the SNR (Stephenson et al. 1977, SG02).

Arabic scholars -- following Aristotle -- considered transient celestial events 
(including comets as well as variable and new stars)
as being located in the Earth atmosphere like meteors, because Aristotle considered true stars to be eternal and constant.
In his report, $^{c}$Al\={\i} ibn Ri\d{d}w\={a}n used the Arabic words {\em athar} and {\em nayzak},
which can both mean something like {\em spectacle}, e.g. a very bright star (Goldstein 1965).
The word {\em athar} can also mean {\em trace}, possibly something like a persistence effect in the eye
due to the strong brightness and/or strong scintillation.
There is also one Arabic report about SN 1054 (Brecher et al. 1978):
Ibn Ab\={\i} U\d{s}aybi$^{\rm c}$a (historian who lived AD 1203-1270 in Damascus, Syria)
quoting Ibn Bu\d{t}l\={a}n (a physician, who lived AD 1038-1075 in Baghdad, Iraq)
wrote about the {\em athar\={\i} kawkab} as
{\em the star leaving traces} or {\em spectacle star} (Brecher et al. 1978);
{\em kawkab} can mean {\em star} or {\em celestial object} in a general sense including {\em planet}
(while {\em najm} can only mean {\em star}).
Regarding the translation of {\em athar}, Brecher et al. (1978) write:
{\em ... a novel astronomical or meteorological phenomenon ... characterised by its
transient, explosive or spectacular appearance. Apparent star, phenomenon or
spectacle might be equally viable translations}.
The word {\em nayzak} was previously and otherwise used for comets, 
as they also appear to be new and transient celestial objects,
so that the {\em new star of 1006} was long regarded as comet instead of a SN (Goldstein 1965).
See Kunitzsch (1995) for a review on the Arabic words used for stars and transient celestial objects.

While the other Arabic and east Asian observations of SN 1006 all were obtained between 
the geographic latitudes of $30^{\circ}$ and $35^{\circ}$ north, 
we discuss here an additional observation from \d{S}an$^{c}$\={a}$^{\prime}$, 
the capital of Yemen, i.e. at $15.3^{\circ}$ north. 
We present for the first time the Arabic text plus a translation to English from al-Yam\={a}n{\={\i}} and then,
also for the first time, the original Arabic text from Ibn al-Dayba$^{c}$ (Sect. 2), both with text variants.
Then, we date the observation by converting from the Islamic calendar to the Julian calendar (Sect. 3).
Finally, we discuss briefly the other information contained in the Yemeni reports (Sect. 4),
and conclude with a summary (Sect. 5).
An English translation of one of the two texts was first presented in SG02 quoting private communication with one of us (WR).

\section{Arabic text about SN 1006 from Yemen}

The two original Arabic texts are shown in Figs. 1-3. 

The first (earlier) text is a small excerpt from the book entitled 
{\em Bahjat al-Zaman f{\={\i}} t\={a}r{\={\i}}kh al-Yaman} 
written by Ab\={u} Mu\d{h}ammad (Ab\={u} l-Ma\d{h}\={a}sin) $^{c}$Abdalb\={a}q{\={\i}} b. $^{c}$Abdalmaj{\={\i}}d b.
$^{c}$Abdall\={a}h T\={a}jadd{\={\i}}n b. Abi l-Ma$^{c}$\={a}l{\={\i}} Muthann\={a} b. A\d{h}mad b. Mu\d{h}ammad b. $^{c}$Is\={a}
b. Y\={u}suf al-Yam\={a}n{\={\i}} al-Makhz\={u}m{\={\i}} al-Qurashi al-$^{c}$Adan{\={\i}} al-Sh\={a}fi$^{c}$i 
(for short: al-Yam\={a}n{\={\i}}),
he died in AD 1342. We use the edition of al-Hubaishi \& al-Sanab\={a}ni (1988), see Fig. 1 for the text about SN 1006.

\begin{figure*}
{\includegraphics[angle=0,width=17cm]{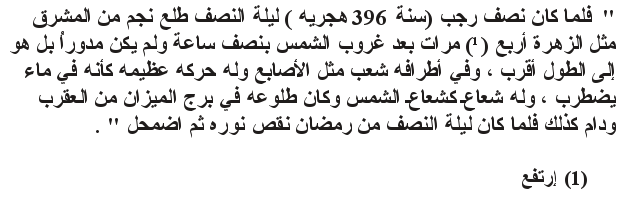}}
\caption{The Arabic text from the History of Yemen from al-Yam\={a}n{\={\i}}, 
here the report about the year 396h including SN 1006.
This is the earlier text no. 1. 
We also show one small text variant (1), bottom line, 
as specified in the edition of al-Hubaishi \& al-Sanab\={a}ni (1988),
page 63:
The text variant says {\em irtafa$^{c}$} instead of {\em arba$^{c}$}.
These two words are written in a very similar way in Arabic.
With the variant {\em irtafa$^{c}$} (Arabic verb for {\em to rise}), 
the text means {\em It was as large as Venus and rose several times after sunset} --
while with {\em arba$^{c}$} (Arabic word for {\em four}), 
it would mean {\em It was four times as large as Venus}.
The version with {\em arba$^{c}$} is better Arabic,
while the version with {\em irtafa$^{c}$} would lead to some kind of a duplication:
{\em ... a star appeared from the east at half an hour after sunset.
It was as large as Venus and rose several times after sunset.}
The later version of the text by Ibn al-Dayba$^{c}$ has just {\em irtafa$^{c}$}.}
\end{figure*}

Our English translation of this text is as follows
(with a text variant given in square brackets, see Fig. 1): \\
{\em On the night of mid-Rajab (or: 15th of Rajab), in the year 396h,
a star appeared from the east at half an hour after sunset.
It was four times as large as Venus.
[It was as large as Venus and rose several times after sunset.]
It was not circular, but nearer to an oblong.
At its ends, there were lines like fingers.
It showed a great turbulence as though it was seen in disturbed water.
Its light rays were similar to sunlight.
It appeared in the zodiacal sign of Libra in Scorpio
and remained unchanged like that.
In the night of mid-Rama\d{d}\={a}n, its light started to decrease and gradually faded away.}

The 2nd (later) text is a small excerpt from the book entitled
{\em Kit\={a}b Qurrat al-$^{c}$uy\={u}n f\={\i} akhb\={a}r al-Yaman al-maim\={u}n} about the history of Yemen,
written by Ab\={u} $^{c}$Abdall\={a}h $^{c}$Abdalra\d{h}m\={a}n ibn $^{c}$Al\={\i} ibn Mu\d{h}ammad ibn $^{c}$Umar
ibn $^{c}$Al\={\i} ibn Y\={u}suf Wagjihald\={\i}n al-Shayb\={a}n\={\i} al-Zab\={\i}d\={\i} ibn al-Dayba$^{c}$
(for short: Ibn al-Dayba$^{c}$),
who was born on AD 1461 Oct 8 in Zah\={\i}b, Yemen,
worked as a teacher and chronicler at the main mosque of Zah\={\i}b, Yemen,
and died on AD 1537 Dec 21, also in Zah\={\i}b, Yemen,
i.e. he lived and wrote about five centuries after SN 1006 (see Brockelmann 1949, Sellheim 1976,
the latter being a translation of the auto-biography of Ibn al-Dayba$^{c}$).

We used manuscript number 416 from the Wadod Center for indexing and edited books;
this manuscipt is a copy written in AD 1680; see Figs. 2 \& 3 for the text about SN 1006. 
There is also an edition of Ibn al-Dayba$^{c}$'s work Qurrat al-$^{c}$uy\={u}n by al-Akwaa' al-Hiw\={a}li 
as publication of the Hiw\={a}li Yamani Library (\d{S}an$^{c}$\={a}$^{\prime}$, Yemen);
the two texts are almost identical -- with one exception: 
For the text by Ibn al-Dayba$^{c}$, in our manuscript (Fig. 2) the word {\em shu$^{c}$b} is missing,
which is present in both the printed edition of the work of Ibn al-Dayba$^{c}$ by al-Akwaa' al-Hiw\={a}li (Fig 3)
and also in the printed edition of the work of al-Yam\={a}n{\={\i}} by al-Hubaishi \& al-Sanab\={a}ni (1988), Fig. 1.
The word {\em shu$^{c}$b} means {\em branches} or {\em lines}; 
the meaning of the text does not change significantly with or without this word.

An English translation of this text is as follows (as in SG02);
the translation of the word {\em shu$^{c}$b} (for {\em lines}) from the edition of al-Akwaa' al-Hiw\={a}li (Fig. 3)
is given in square brackets below, it is missing in our manuscript (Fig. 2): \\
{\em In the year 396h, on the night of mid-Rajab, 
a star like Venus appeared.
It regularly rose half an hour after sunset.
It was not round, but rather was elongated;
at its edges were [lines like] fingers.
It showed great turbulence as though (reflected) in disturbed water.
Its light rays were similar to the rays of the Sun.
It appeared in the location of al-Ghafr in the sign of Libra.
It remained unchanged until the night of mid-Rama\d{d}\={a}n.
Then, its light deminished and it gradually faded away.} 

In manuscript number 416 as consulted by us, see Fig. 2, there is also a comment
or headline about this new star at the left margin saying (Fig. 2): \\
{\em the appearance of a great ranking star} \\
which may have been written by the original author or a copying scribe.

Neither of the two texts was written by the observer, but they are histories written several hundred years after the SN.
For both texts, we have to assume that the manuscripts known today are copies of copies of copies etc.
The fact that they are otherwise very similar shows that the later text (written between AD 1461 and 1537)
is probably derived from the earlier text (written before AD 1342) or its source.

Let us now discuss the differences of the two texts:
\begin{itemize}
\item al-Yam\={a}n{\={\i}}: {\em On the night of mid-Rajab, in the year 396h, a star appeared  ...}, \\
Ibn al-Dayba$^{c}$: {\em In the year 396h, on the night of mid-Rajab, a star like Venus appeared}.
\item al-Yam\={a}n{\={\i}}: {\em It was four times as large as Venus}, \\
Ibn al-Dayba$^{c}$: {\em a star like Venus appeared}, \\
i.e. the latter left out the factor {\em four}
(text variant of al-Yam\={a}n{\={\i}} also without that factor, but then in less good Arabic).
\item al-Yam\={a}n{\={\i}}: {\em a star appeared from the east at half an hour after sunset}, \\
Ibn al-Dayba$^{c}$: {\em It regularly rose half an hour after sunset}, \\
i.e. leaving out the direction and adding {\em regularly}.
\item al-Yam\={a}n{\={\i}}: {\em It appeared in the zodiacal sign of Libra in Scorpio}, \\
Ibn al-Dayba$^{c}$: {\em It appeared in the location of al-Ghafr in the sign of Libra}, \\
i.e. {\em al-Ghafr} instead of {\em Scorpio}.
\item al-Yam\={a}n{\={\i}}: {\em It appeared ... and remained unchanged like that. 
In the night of mid-Rama\d{d}\={a}n, its light started to decrease and gradually faded away}, \\
Ibn al-Dayba$^{c}$: {\em It remained unchanged until the night of mid-Rama\d{d}\={a}n. 
Then, its light diminished and it gradually faded away}, \\
i.e. Ibn al-Dayba$^{c}$ combined the last sentences from al-Yam\={a}n{\={\i}},
but both texts can be interpreted to show stationarity. 
\end{itemize}
The later text by Ibn al-Dayba$^{c}$ is shorter and/or less precise than the earlier text from al-Yam\={a}n{\={\i}},
e.g. Ibn al-Dayba$^{c}$ left out the factor {\rm four} in the brightness. 

Regarding the difference in size or brightness in the two Yemeni reports, it may seem likely at first sight,
that the 2nd author just left out the words {\em four times} (as large as Venus) and wrote instead {\em a star like Venus}.
One could consider that he left out {\em four times}, because, e.g., it may have seem to him to be too much or too bright,
but otherwise he left in {\em its light rays were similar to the rays of the Sun}, also appearing to be very bright.
It is not certain that the 2nd author copied directly from the other (earlier) manuscript that we know today.
It is well possible that he copied from an earlier, possibly different source, 
from which also the 1st report was copied earlier.
The difference in this regard appears only in one of the two text variants. \\
Therefore, it seems possible that the 2nd author, Ibn al-Dayba$^{c}$, did not copy from the text in the very version
we call here text no. 1, but that he either copied from an even earlier version of that report,
or that he changed some details (e.g. {\em as Venus}, 
instead of {\em four times as large}), because he thought to know better from
yet another source. However, this possibility appears more speculative.
Of course, we cannot exclude that the differences are due to 
(intentional or unintentional) changes or mistakes made by copying scribes.

The main variant of the first text from al-Yam\={a}n{\={\i}} is better Arabic than its minor variant and
than the text from Ibn al-Dayba$^{c}$,
and it is also more detailed,
so that we conclude that the 2nd text is most certainly derived from the first one.

Note that in this text about a new star (SN 1006), the word {\em najm} is used for {\em star},
while in the previously found Arabic reports about SN 1006 and 1054 the words {\em kawkab} (used for star or planet),
{\em nayzak} (spectacle), and {\em athar} are used ({\em athar}\textit{\={\i}} {\it kawkab} meaning
{\em the star leaving traces} or alternatively {\em spectacle star}, Brecher et al. 1978).

The reports from Yemen offer new and independant information on SN 1006,
which is different from all other known reports,
e.g. the apparent early discovery date, the detection half an hour after sunset, and the brightness (4 times as Venus).
Hence, the reports are probably original (hence, likely from Yemen), and not copies of reports from other countries.

In the following interpretation, we will of course consider both reports, their differences, and the different text variants.

\begin{figure*}
{\includegraphics[angle=0,width=17cm]{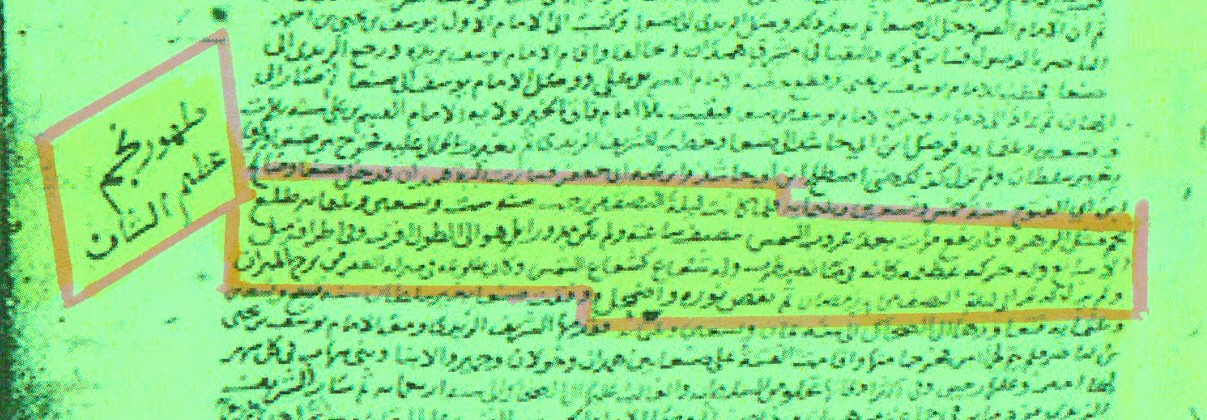}}
\caption{A copy of a small part of the History of Yemen from Ibn al-Dayba$^{c}$,
here the report about the year 396h including SN 1006.
This is the younger text no. 2.
It is taken from manuscript number 416 (written AD 1680) from the Wadod Center for indexing and edited books.
An edition of Ibn al-Dayba$^{c}$'s work Qurrat al-$^{c}$uy\={u}n by al-Akwaa' al-Hiw\={a}li is available
as publication of the Hiw\={a}li Yamani Library (\d{S}an$^{c}$\={a}$^{\prime}$, Yemen).
The relevant text about SN 1006 is indicated by yellow background:
It starts before the middle of the 7th line from the bottom with
{\em Fa-lamm\={a} k\={a}nat laylat al-ni\d{s}f min Rajab ... (on the night of mid Rajab ...)};
it ends after the middle of the 4th-to-last line with
{\em ... thumma naqasa n\={u}ruhu wa-i\d{d}ma\d{h}alla (... then its light diminished and it gradually faded away.)}.
The words at the left margin say {\em \d{z}uh\={u}r najm $^{c}$\d{z}{\={\i}}m al-sha`n 
(the appearance of a great ranking star)}.
At the end of the 6th line from the bottom, the last three words read {\em waf\={\i} a\d{t}r\={a}fihi mithlu},
while the same portion in the text of al-Yam\={a}n{\={\i}}, it says {\em waf\={\i} a\d{t}r\={a}fihi shu$^{c}$b mithlu};
the extra word {\em shu$^{c}$b} means {\em branches} or {\em lines}, 
so that the meaning of these two versions is not really different.
See Fig. 3 for the Arabic text re-written by us for better reading. Our translation is given in Sect. 2.
}
\end{figure*}

\begin{figure*}
{\includegraphics[angle=0,width=17cm]{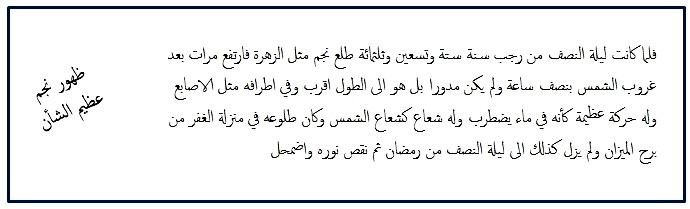}}
\caption{The Arabic text from the {\em Qurrat al-$^{c}$uy\={u}n} by Ibn al-Dayba$^{c}$, here the report about the year 396h including SN 1006,
re-written by us from the edition of al-Akwaa' al-Hiw\={a}li. Please note that the manuscript (Fig. 2) does not have the word
{\em shu$^{c}$b}, which is present in the edition by al-Akwaa' al-Hiw\={a}li.}
\end{figure*}

\section{Dating} 

According to de Blois (2000), the wording {\em ni\d{s}fi min Rajab} means {\em middle of} or 
{\em 15th of Rajab}, translated as {\em mid-Rajab} in SG02.
This wording is not unusual for Arabic, it does mean one particular night (or day) 
and is not a rough indication of the middle of the month.
The dates given in both Yemeni reports ({\em the night of mid-Rajab} or 15th of Rajab) 
and {\em the night of mid-Rama\d{d}\={a}n} or 15th of Rama\d{d}\={a}n in the year 396h) 
will now be converted from the Islamic calendar to the Julian calendar.

The {\em calculated} Islamic calendar can be converted to the Julian (or Gregorian) 
calendar by starting the Islamic calendar with 
year 1h on AD 622 Jul 15/16,\footnote{The Islamic year 396 hijra (396h) started 396 
lunar years after the start of the lunar year in which the Hijra took place, 
i.e. the emigration of the Islamic Prophet Mu\d{h}ammad from Mecca to Medina, known as Hijra; 
this era, i.e. the year 1h started on AD 622 Jul 16 (evening of 15th to evening of 16th July)
according to most scholars -- but it may have been on AD 622 Jul 15 (evening of 14th to evening of 15th July)
see, e.g., de Blois (2000); according to NASA GSFC (eclipse.gsfc.nasa.gov/phases/), 
new moon was on AD 622 Jul 14 (Julian calendar)
at 5:26h UT ($\pm 2$ min, Morrison \& Stephenson 2004), so that the crescent new moon was hardly visible
in the evening of AD 622 Jul 14, but it was well visible in Mecca, 
Saudi Arabia, in the evening of AD 622 Jul 15 (Neugebauer 1929).}
then alternating months with 29 or 30 days, and also one leap day in 11 of every 30 years
(the synodic month is 29.53 days on average).
This is the {\em calculated} Islamic calendar, which may well be different from the real calendar used.
In particular, the length of the months were not set to be alternating between 29 and 30 days,
but the start (and end) of a month was set by observing the crescent new moon.
There are two uncertainties in this {\em calculated} Islamic calendar:
(i) The start of year 1h could have been AD 622 Jul 15/16 or 14/15 (see footnote 1) and 
(ii) it is not clear a-posteriori when in history a month had an extra 
day\footnote{Given that the average duration of a synodic month is 29.53 days,
one needs one extra day in 11 of each 30 years; the calculated Islamic calendar uses leap days in
certain, pre-defined years and months; in reality, we have to expect that, in each period of 30 years, 
there were 11 months which had an extra day (due to real crescent sighting) -- 
in addition to those 354 days in a pure lunar calendar with -- on average -- six months of 29 days and  -- on average -- six months of 30 days.
Due to crescent sighting, the month with one extra day did not neccessarily follow the leap day/month rule in the
calculated Islamic calendar used in, e.g., Spuler \& Mayr (1961).}.
Furthermore, the calulated Islamic calendar was not used in 
history.\footnote{al-B\={\i}r\={u}n\={\i} (AD 973-1048) 
wrote: {\em the leap years of the Arabs, not that the Arabs ever actually used or use them, but the authors of astronomical tables need them when they
construct tables on the basis of the years of the Arabs} (quoted after de Blois 2000)}
In summary, the calculated Islamic calendar can deviate by up to two days (Ginzel 1906, Spuler \& Mayr 1961, de Blois 2000). 

The start (and end) of a month was not set by some calculated calendar, but by real observations of the
crescent new moon: A month started on the evening, when a new crescent moon was seen (Quran, Sura 2, 189).

However, the sighting of a crescent new moon could have been delayed by one or more days due to,
e.g., bad weather, high extinction at low latitude, or difficult landscape.
It may also be possible that a month started one day too early due to a false early 
detection (Doggett \& Schaefer 1994, Ilyas 1994).
If the weather was overcast for several days, the start of the month can be set later
given the age or phase of the moon, also typicaly to $\pm 2$ days.
In some cases, however, it is possible to convert from the Islamic calendar to the Julian (or Gregorian) calendar 
with an accuracy better than $\pm 2$ days
(i.e. better than with the calculated Islamic calendar), namely when some calibration is known,
e.g. the mentioning of an Islamic date together with a week-day, or the observation of a certain astronomical
event, that can be dated nowadays with higher accuracy. 
If the length of a month was fixed to at least 29 days
and/or at most 30 days (which is likely, but not certain, de Blois 2000), then a calibration in the next or
preceeding month might also be sufficient for an accuracy of better than $\pm 2$ days. 
See, e.g., Spuler \& Mayr (1961), Spuler (1963), Ilyas (1994), de Blois (2000), Said et al. (1989), and Neuh\"auser \& Kunitzsch (2014)
for more details about Muslim calendar rules and calendar conversion. 

By converting the date given as {\em night of mid-Rajab} or {\em night of 15th of Rajab} 
from the calculated Islamic calendar (Spuler \& Mayr 1961) to the Julian calendar, we obtain the night of AD 1006 Apr 17/18
(first detection of SN 1006).

Let us now try to determine the date of the first crescent visibility with a precision of better than $\pm 2$ days.

Start of the month Rajab: 
According to eclipse.gsfc.nas a.gov and Gautschy (2011),
new moon (true conjunction of moon and Sun) was on AD 1006 Mar 31 at 23:57h UT ($\pm$ few min, Morrison \& Stephenson 2004),
or on Apr 1 at 02:57h in the current time zone of Yemen (3h east of Greenwich),
or on Apr 1 at 02:54h at the eastern longitude of \d{S}an$^{c}$\={a}$^{\prime}$ ($44^{\circ} 12^{\prime} 23^{\prime \prime}$E).
According to Gautschy (2011), the crescent new moon was 
then visible not before 16:14h UT (in Babylon, Iraq) on AD 1006 Apr 2
(the crescent can be detected only if the separation between sunset and moonset 
is sufficient and if the moon is sufficiently high above horizon, see, e.g., Ilyas (1994) and references therein).
The crescent probably has been sighted at \d{S}an$^{c}$\={a}$^{\prime}$ first in the evening 
of (our) Apr 2 ($19^{\circ}$ above horizon at sunset).
In case of bad weather, an even later sighting is possible.
In case of an early (false) detection, the first sighting of the crescent
could have been reported (and accepted) for Apr 1.

Given that dates are memorized with respect to the end of a month (de Blois 2000), 
we also have to consider the end of the month of Rajab:
According to eclipse.gsfc.nasa.gov and Gautschy (2011),
new moon was on AD 1006 Apr 30 at 9:08h UT ($\pm$ few min, Morrison \& Stephen\-son 2004),
or at 12:08h noon in the current time zone of Yemen, or 12:05h at \d{S}an$^{c}$\={a}$^{\prime}$.
According to Gautschy (2011), the crescent new moon was then visible not before 16:21h UT (for 
Babylon, Iraq) on AD 1006 May 1.
Hence, the first detection of the crescent probably happened on the evening of (our) May 1.

The first day of the new lunar month of Sha$^{c}$b\={a}n would then run from the evening of May 1 to the evening of May 2 
(or maybe from May 2 to 3 in case of bad weather on May 1);
in the less likely case of a false early detection of a crescent, the month could have started on Apr 30.
While according to de Blois (2000) a wording like mid-Rajab means the 15th of Rajab,
all dates since the 15th were 
given and memorized under the {\em assumption} that the month would have a total of 30 days.
One could question this by arguing that, after it was noticed that a particular month had only 29 days,
one could correct the dating; on the other hand, experienced observers of the moon can deduce the number of remaining days
before a new crescent moon could be observable from the currently observed 
age or phase
of the moon anytime during the month.
As we will see below due to sightings of SN 1006, the month of Rajab in that year most certainly had 30 days.
Furthermore, it may be easy to correctly give the date as {\em middle} of the month, because this is near full moon,
which is easy to observe and to determine; according to eclipse.gsfc.nasa.gov, full moon was on AD 1006 Apr 16 at 9:35h UT.
Note, however, that in a lunar month starting with the first sighting of the crescent (and not with true conjunction),
full moon is not exactly in the middle of the month.

Unfortunately, neither al-Yam\={a}n{\={\i}} nor Ibn al-Dayba$^{c}$ mention the weekday of the 15th of Rajab.
We can, however, consider other Arabic reports about SN 1006 for a more precise dating of the end of Rajab.

Ibn al-Jawz{\={\i} reported that SN 1006 was first detected on {\em Friday} at the beginning of Sha$^{c}$b\={a}n --
meaning the evening at the start of the 1st of Sha$^{c}$b\={a}n just after sunset.
And Ya\d{h}y\={a} ibn Sa$^{c}$\={\i}d al-An\d{t}\={a}k\={\i} reported that it was first detected on Saturday, 
in the evening at the start of the 2nd day of Sha$^{c}$b\={a}n.
Hence, for those two additional Arabic observers, the month of Sha$^{c}$b\={a}n started on a Friday (1st of Sha$^{c}$b\={a}n).
In the Islamic calendar, a day (like this {\em Friday}) started at an evening sunset and ended at the next evening sunset
(on the Julian or Gregorian calendar, 
we regard the time from the first evening sunset to midnight as the previous day, here a Thursday).
According to Spuler \& Mayr (1961), our May 3 was indeed a Friday in AD 1006.
Hence, we can identify the 1st of Sha$^{c}$b\={a}n on the Islamic calendar with the date AD 1006 May 2/3 
(from sunset to sunset) on the Julian calendar.
The reports by Ibn al-Jawz{\={\i} and Ya\d{h}y\={a} ibn Sa$^{c}$\={\i}d al-An\d{t}\={a}k\={\i} are consistent with each other,
i.e. the observers, on whose observations the reports are based, both started the month of Sha$^{c}$b\={a}n at the
evening sunset of AD 1006 May 2. The observer, whose observations are reported by Ibn al-Jawz{\={\i}, detected
both the lunar crescent and SN 1006 for the first time on that evening (hence, 1st of Sha$^{c}$b\={a}n).
For the observer, whose observations are reported by Ya\d{h}y\={a} ibn Sa$^{c}$\={\i}d al-An\d{t}\={a}k\={\i},
the weather was also clear and fine on the evening of May 2, so that he could detect the crescent --
but he did not detect the supernova on that evening. \\
Note that the start of the lunar month on the evening of AD 1006 May 2 in Arabia is not in contradiction
with the Far Eastern reports about SN 1006, where we can read:
{\em on the 2nd day of the 4th lunar month (May 1)} (China, Song Huijao Jigao) and
{\em on the 2nd (day) of the 4th month (May 1)} (Japan, Meigetsuki) 
(texts cited and date conversions given are both from SG02),
i.e. the lunar month started in China and Japan on Apr 30 (days and dates run from midnight to midnight).
As mentioned above, new moon (conjunction of moon and sun) was on AD 1006 Apr 30 at 9:08h UT.
And indeed, the Chinese started the day-count in each (lunar) month with what we call {\em new moon}, 
i.e. conjunction of moon and sun, as confirmed by the fact that all of the dates of 
solar eclipses from (at least) AD 700 to 1200 are dated to {\em the first day of the month},
see listing in Xu et al. (2000). \\
According to the calculated Islamic calendar, the 1st of Sha$^{c}$b\={a}n (from evening to evening) was on AD 1006 May 2/3,
which is here consistent with the true crescent detection by the Arabic observers.
According to Gautschy (2011), the crescent would have been visible on AD 1006 May 1, but we have to conclude that 
the weather may have been bad on that evening, so that the month started one day later (and SN 1006 was also not detected
by those observers on May 1 due to bad weather). Since the month could have started one day earlier than it did,
the previous month probably had 30 (instead of 29) days. \\
$^{c}$Al\={\i} ibn Ri\d{d}w\={a}n observed SN 1006 on AD 1006 Apr 30, without giving excplicitely this date,
but this dating (and the fact that he had good weather on that date) is not inconsistent with the weather
of the other observers discussed above on May 1 and 2.

Let us now assume that the weather was similar (good on May 2 and bad on May 1 evening) for the observer, whose observations are reported by 
al-Yam\={a}n{\={\i}} and Ibn al-Dayba$^{c}$, 
so that he also started the month of Sha$^{c}$b\={a}n at the evening sunset of AD 1006 May 2.

If the 15th of Rajab was given with respect to the end of the month under the assumption that it had 30 days
(see above, de Blois 2000), then the 15th of Rajab ran from the sunset of AD 1006 Apr 16 to the next sunset on Apr 17.

Above, we had concluded that the possible dates for the evening of the 1st of Rajab were 1 Apr (early false detection)
or Apr 2 (best case) or Apr 3 (in case of bad weather on Apr 2).
If the month of Rajab ended with the evening sunset of AD 1006 May 2, 
and if its length was 30 days, then it started on Apr 2 at evening sunset
(if its length was 29 days, then it started on Apr 3 at evening sunset).

An additional uncertainty of one more day could be introduced, if a month would last neither 29 nor 30 days;
while some scholars argue that this is excluded given an Hadith\footnote{According
to the collection of Hadith by Bukh\={a}r\={\i} translated by M. Muhsin Khan (www.searchtruth.com/hadith$\_$books.php),
Prophet Mu\d{h}ammad once said: {\em When you see the crescent (of the month of Rama\d{d}\={a}n),
start fasting, and when you see the crescent (of the month of Shaww\={a}l), stop fasting;
and if the sky is overcast (and you can't see it) then regard the crescent (month) of Rama\d{d}\={a}n (as of 30 days).}},
others argue that this Hadith applies only to the month of Rama\d{d}\={a}n and could show
evidence that in a few rare cases, a Muslim month did last 31 days (see, e.g., de Blois 2000, according to
Schaefer (1993), this can happen in $0.06~\%$ of cases).
The most likely dates for the start of the night of the 15th of Rajab remain AD 1006 Apr 16 and 17.

In the evenings of AD 1006 Apr 17 and 18, the moon was below the horizon for a significant time
after the rising of SN 1006, so that it should have been possible to have detected SN 1006 since AD 1006 Apr 17,
while on the evening of AD 1006 Apr 16, the moon was above the horizon (and also above the location of SN 1006)
before SN 1006 rose, so that it would have been less likely to discover it.
If the weather was not good enough for the detection of the crescent of AD 1006 Apr 2 at the site of the observer, 
whose observations are reported by al-Yam\={a}n{\={\i}} and Ibn al-Dayba$^{c}$, but good enough on Apr 3, 
then the month of Rajab started (for him) on Apr 3, so that on Apr 17 (evening) the 15th of Rajab started.

Al-Yam\={a}n{\={\i}} then also mentioned that SN 1006 {\em a star appeared from the east at half an hour after sunset}
(Ibn al-Dayba$^{c}$: {\it regularly rose half an hour after sunset}):
At the location of \d{S}an$^{c}$\={a}$^{\prime}$ in Yemen, the apparent rising of SN 1006 
(on an assumed perfectly flat horizon) was indeed
$\sim 25$ to 30 min after apparent sunset around AD 1006 Apr 17. This time difference then decreased until SN 1006 rose
at sunset around AD 1006 Apr 23/24. After that date, SN 1006 rose before sunset.  \\
This conclusion is correct for considering {\em half an hour} either as a period 
of 30 well-defined clock-minutes (as we do today)
or as a period of half a sun-dial hour (as done in former times, when the bright day-light time was split
into 12 {\em hours} of unequal duration), because in mid-April, 
half a sun-dial hour was only slightly longer than 30 clock-minutes,
in particular in a location as south as Yemen. \\
The considered rise time is also consistent with another historic report, the Mauretanian report said for early May: \\
{\em (It appeared on the first (day) of Sha$^{c}$b\={a}n, (3)96 mentioned above.)
Its first appearance was before sunset, whereupon it faded until night came and it reappeared} \\
(SG02), but translated different in Goldstein (1965): \\
{\em It began to appear in the beginning of Sha$^{c}$b\={a}n (3)96.  
Its first appearance took place before setting whereupon it went backwards until it rose night.}\footnote{Both translations 
specify that SN 1006 was detected before sunset, i.e. during the late day.
There is no such day-time detection reported from the Chinese, Japanese, nor Koreans.
The day-time detection from Mauretania may have been facilitated by clean air and/or high altitude
(the hightest point in today's Mauretania is the Kediet al-Jill at 915 m,
and the highest point in today's Morocco is the Atlas Mountain with 4,165 m in the north of Morocco).
Since the conversion of a Muslim date to a Julian date depends on whether SN 1006 was detected before or after sunset,
i.e. before or after the change of the date, we have to clarify the date here:
The reported detection on the {\em first (day) of Sha$^{c}$b\={a}n} (or {\em the beginning of Sha$^{c}$b\={a}n})
before sunset would then not neccessarily be the evening of AD 1006 May 2, but maybe already May 3 just before sunset,
because the first day of Sha$^{c}$b\={a}n ran from the evening of AD 1006 May 2 to the evening of May 3.
This consideration would modify the detection date from Mauretania (as given in Goldstein 1965 and SG02) by one day.
Detection of SN 1006 before sunset would have been possible from Mauretania/Morocco best since AD 1006 May 5.}

The fact that the difference between rise time and sunset is fully consistent 
with the given discovery date (around AD 1006 Apr 17 evening),
for a location of \d{S}an$^{c}$\={a}$^{\prime}$ in Yemen,
gives quite a high confidence and credibility in the report by al-Yam\={a}n{\={\i}}.
It can also be seen as circumstancial evidence that the observations were really done at (or near) \d{S}an$^{c}$\={a}$^{\prime}$ in Yemen.

It may be quite surprising if the observer, who is the original source for the reports of the Yemeni authors,
really has observed and detected SN 1006 already several days before the other Arabic observers 
(e.g. Apr 30 by $^{c}$Al\={\i} ibn Ri\d{d}w\={a}n), and all (confirmed) eastern Asian observers (not before May 1),
so that this early date was rejected by SG02 as artificial.

However, since \d{S}an$^{c}$\={a}$^{\prime}$ is at 2400 m sea level and since it has a clear horizon towards the south,\footnote{The Jabal al-Nabi Shu$^{c}$ayb,
the highest mountain in Arabia with 3666 m, is due west-south-west 
from \d{S}an$^{c}$\={a}$^{\prime}$, while SN 1006 was rising in the
evening in the south-east.} an observation from here earlier than all other known detections may not appear impossible.

\section{Brightness, duration of visibility, lightcurve, location, stationarity, and color} 

We will now discuss the other information content from the Yemeni reports.

{\bf Brightness.}
Al-Yam\={a}n{\={\i}} gives {\em it was four times as large as Venus} (in the more likely text variant),
while Ibn al-Dayba$^{c}$ gives {\em a star like Venus}.
If we interprete this observation as brightness of the new star, than it was (probably) four times brighter than Venus,
i.e. 1.5 mag brighter than Venus around that time.
Venus was visible in the west increasing in apparent brightness from $-2$ to $-3$ mag;
in Sep 1006, Venus and SN 1006 came as close as $16^{\circ}$ with Venus having an apparent magnitude $-4$ mag.
Alternatively, {\em four times as large as Venus} could mean the apparent size instead of the brightness.

{\bf Duration of visibility.}
The last sentence of al-Yam\={a}n{\={\i}} is: 
{\em In the night of mid-Rama\d{d}\={a}n, its light started to decrease and gradually faded away}, 
similar in Ibn al-Dayba$^{c}$. 
According to the calculated Islamic calendar (Spuler \& Mayr 1961),
the {\em night of mid-Rama\d{d}\={a}n} (or 15th of Rama\d{d}\={a}n) is AD 1006 Jun 15/16.
According to eclipse.gsfc.nasa.gov, conjunction between moon and Sun was on AD 1006 May 29 at 19:32h UT,
and according to Gautschy (2011), the crescent new moon was not visible (at Babylon, Iraq) before AD 1006 May 31 at 16:45 UT.
Hence, the 1st day of Rama\d{d}\={a}n 396h started probably on the evening of AD 1006 May 31,
and the 15th of Rama\d{d}\={a}n was then AD 1006 June 14/15 (evening to evening).
On this day, SN 1006 started to get fainter (in the Yemeni reports), so that the period of visibility is at least two months. \\
According to eclipse.gsfc.nasa.gov, full moon was on AD 1006 June 14 at 7:46h UT 
(with a partial lunar eclipse visible in Arabia),
so that it was relatively easy to date the middle of Rama\d{d}\={a}n well. Maybe the bright light of
the moon in and around that night caused the observer to conclude that SN 1006 was getting fainter. \\
The last sentences of al-Yam\={a}n{\={\i}} ({\em In the night of mid-Rama\d{d}\={a}n, its light started to decrease and gradually faded away}),
and also even in Ibn al-Dayba$^{c}$ ({\em its light diminished and it gradually faded away}), both for mid June,
do not mean that the Yemeni observations were restricted to those two months from mid-April to mid-June,
but rather that the brightness of SN 1006 remained roughly constant until mid-June, and then started to get fainter.
$^{c}$Al\={\i} ibn Ri\d{d}w\={a}n reported: {\em it ceased all of a sudden} in July, i.e. indeed even one month later. 

{\bf Lightcurve.}
While the more likely text variant in al-Yam\={a}n{\={\i}} says that the SN was four times brighter (or larger) than Venus
(in mid April 1006),
the less likely text variant and Ibn al-Dayba$^{c}$ give a different brightness: {\em a star like Venus}. \\
As discussed above, it is not impossible that the (later) report by Ibn al-Dayba$^{c}$ regarding the brightness is correct,
either because he corrected the other report by al-Yam\={a}n{\={\i}} with independant information
(or, less likely, that the two values refer to different dates, even though they report the same dates).
From the reports by Ibn al-Dayba$^{c}$ ({\em a star like Venus}) and $^{c}$Al\={\i} ibn Ri\d{d}w\={a}n 
({\em 2.5 or 3 times as large as Venus}), 
we may be able to conclude that SN 1006 was about 1 mag fainter when observed by the source of Ibn al-Dayba$^{c}$
than as it was when observed by $^{c}$Al\={\i} ibn Ri\d{d}w\={a}n.
Given the typical lightcurve of a SN Ia, it increases in brightness for the first 10 to 18 days (before the peak) by 1 to 1.5 mag,
and it also decreases by 1 to 1.5 mag in 10-18 days after the peak (Woosley et al. 2007, Gameshalingam et al. 2011).
The presumable magnitude and time difference between the two observations reported 
by Ibn al-Dayba$^{c}$ and $^{c}$Al\={\i} ibn Ri\d{d}w\={a}n would then be consistent with a SN Ia lightcurve.
From this consideration, it remains uncertain as to whether the observation reported by Ibn al-Dayba$^{c}$
was 10-18 days before or after the observation reported by $^{c}$Al\={\i} ibn Ri\d{d}w\={a}n.
It would be fully consistent with a SN Ia lightcurve, if the observation reported by Ibn al-Dayba$^{c}$ (e.g. evening of Apr 17)
would have been 10-18 days before the one reported by $^{c}$Al\={\i} ibn Ri\d{d}w\={a}n (Apr 30).
While SN II-P show slower brightness decreases than II-L, the historic data on the lightcurve
of SN 1006 are not sufficient to distinguish between SN types.
Given that al-Yam\={a}n{\={\i}} may report a different brightness for the same date as Ibn al-Dayba$^{c}$, 
the above consideration is highly speculative,
even though this difference appears only in one of the two text variants of al-Yam\={a}n{\={\i}}.
Furthermore, also in the report by $^{c}$Al\={\i} ibn Ri\d{d}w\={a}n, the wording {\em 2.5 or 3 times as large as Venus}
can be interpreted as meaning the apparent size instead of the brightness.

{\bf Location.}
Al-Yam\={a}n{\={\i}} gives the position as {\em in the zodiacal sign of Libra in Scorpio},
while Ibn al-Dayba$^{c}$ gives {\em in the location of al-Ghafr in the sign of Libra}.
The reports could either refer to the ecliptic longitude of the new star by comparing it to 
the ecliptic longitude of a constellation or zodical sign (like Libra and Scorpio) or a lunar mansion (al-Ghafr),
or the reports could give the location of the new star (e.g. meaning in Libra or in Scorpio or in al-Ghafr).
The wording {\em Libra in Scorpio} can be interpreted in different ways: \\
Roughly the border 
between the two (astrological) zodiacal signs, \\
roughly as the area of overlap of the constellations Libra and Scorpio, \\
in the stellar constellation of Scorpio, but also in the zodical sign of Libra, \\
or the area in Libra with Arabic lunar mansions, that pertain to (and are named after) the constellation of Scorpio,
i.e. roughly in the 2nd half of Libra. \\
In historic Arabic astronomy (and astrology), {\em al-Ghafr} ({\em the covering} 
or {\em the hair on the end of the tail of the beast of prey})\footnote{sometimes
also given as {\em the hair on the end of the lion's tail}, which is, however, misleading, 
because it is not connected to today's
constellation of {\em Lion} (Leo), but to the Arabic constellation/asterism of {\em beast of prey}.}
is the 15th of 28 lunar mansions (as already given in SG02)
running from $0^{\circ}$ to $12^{\circ}51^{\prime}26^{\prime \prime}$ within the zodiacal sign of Libra.
There are 28 lunar mansions around the ecliptic (one sideral month lasts 27.3 days).
The lunar mansions with their names and borders can be found, e.g., in {\em The Chronology of Ancient Nations}
by the famous Arab astronomer {\it al-B\={\i}r\={u}n\={\i}} (AD 973-1048), edited and translated to English by Sachau (1879),
as well as in the book called Gh\={a}yat al-\d{h}ak\={\i}m (Latin Picatrix) about astrology written ar\-ound AD 1000
(Hartner 1965, Sezgin 1971, Pingree 1986). \\
Depending on the interpretation, the location given for SN 1006 can be correct: 
It has an ecliptic longitude as near the border of the true stellar constellations of Libra and Scorpio 
as seen around AD 1006, and SN 1006 also has such an ecliptic longitude.

{\bf Stationarity.}
Al-Yam\={a}n{\={\i}} wrote: {\em It appeared in the zodiacal sign of Libra in Scorpio and remained unchanged like that.}.
This can be interpreted such as that the object {\em remained unchanged} with respect to the stars,
i.e. that it did not move with the planets, but with the stars, i.e. that it was stationary. \\
The Mauretanian report said for early May that SN 1006 moved {\em backwards} and
$^{c}$Al\={\i} ibn Ri\d{d}w\={a}n reported: {\em It remained where it was and it moved daily with its zodiacal sign} --
probably both meaning that it was staying still relative to the stars.

{\bf Color.}
If the text given by Ibn al-Dayba$^{c}$ ({\em a star like Venus}) 
would refer to the color of the SN (instead of the brightness),
it would have a Venus-like color, i.e. white to yellow. However, the text by Ibn al-Dayba$^{c}$ is probably
just a shortened version of the text by al-Yam\={a}n{\={\i}}, who clearly said: {\em it was four times as large as Venus}.

\section{Summary}

We have presented the Arabic texts of the observation of SN 1006 by al-Yam\={a}n{\={\i}} and Ibn al-Dayba$^{c}$ from Yemen.
While the English translation of the text from Ibn al-Dayba$^{c}$ was presented earlier (SG02), 
the more original text from al-Yam\={a}n{\={\i}} is presented here for the first time.
We have also discussed their content in comparison with other Arabic and eastern Asian observations.

Brightness (or size or color), duration of visibility (at least two months), stationarity
(with respect to the stars), and position are fully consistent with the other Arabic and eastern Asian observations,
and of course fully consistent with all what is known about SN 1006, which gives a lot of confidence in the reports.
Yet, the most striking difference is the early discovery around AD 1006 Apr 17.
The very early sighting in \d{S}an$^{c}$\={a}$^{\prime}$, Yemen, might have been facilitated by its
large height above sea-level being some 2400 m: Going from such a large height to roughly sea level,
where most of the other, later observers were located (e.g. Cairo, Japan, China) can change the
atmospheric extinction for object low on the horizon by some 4 mag (e.g. Schaefer 1993).
If the brightness of SN 1006 was around -3 mag (like Venus at that time) when observed at the location of \d{S}an$^{c}$\={a}$^{\prime}$, Yemen,
it would have been $\sim 1$ mag at the other sites, so that it might not have been noticed, yet,
as new bright star -- the limit for serendipitous discovery of a new star on the sky by naked-eye is
some 0 to 2 mag according to Clark \& Stephenson (1977) and Strom (1994).

It might be possible that the author (al-Yam\={a}n{\={\i}}, Ibn al-Dayba$^{c}$, and/or an earlier author) found in his source
only the month, not the date, and that at least one of them thought that he would need to give the date with better
precision, i.e. not only the month, but also the day in the month, he then gave {\em night of mid-Rajab} as his best guess.
(The same consideration can be applied to the last obervational date given ({\em the night of mid-Rama\d{d}n}), as already considered by SG02.)
On the other hand, the wording {\em the night of mid-Rajab} is quite typical and normal for Arabic dates,
and it does mean exactly the 15th of the month (de Blois 2000).
Also, the report that {\em it rose several times half an hour after sunset} is fully consistent with AD 1006 Apr 17 ($\pm 2$ day)
for an observation from Yemen, in particular from as high as its capital \d{S}an$^{c}$\={a}$^{\prime}$.
The otherwise earliest certain observation is on AD 1006 Apr 30 by $^{c}$Al\={\i} ibn Ri\d{d}w\={a}n.

\acknowledgements
We would like to thank Daniel Fischer, thru whom the two authors got into contact with eachother.
We acknowledge the moon phase predictions by Rita Gautschy 
and Fred Espenak, NASA/GSFC, on eclipse.gsfc.nasa.gov.
WR would like to thank Sahi Hassoun al-Ta`i of Hilla University College for obtaining 
the second manuscript from the Wadod Center;
we would like to acknowledge the Wadod Center for indexing and edited books, 
which was established in memory of the female Sheikha al-Murry;
RN also thanks the Institut f\"ur Geschichte der Arabisch-Islamischen Wissenschaften, Frankfurt,
where he consulted the al-Hubaishi \& al-Sanab\={a}ni edition of al-Yam\={a}n{\={\i}}'s 
work Bahjat al-Zaman f\={\i} t\={a}r{\={\i}}kh al-Yaman as well as the auto-biography of Ibn al-Dayba$^{c}$.
RN thanks Hani Dalee and Mazen Amawi for advise on the Arabic text.
RN would also like to thank D.L. Neuh\"auser for advise on the 
dependance of the texts and the interpretation regarding the location of the SN.
RN acknowledges the German national science foundation (Deutsche Forschungsgemeinschaft, DFG)
for financial support in the collaborative research center Sonderforschungsbereich SFB-TR 7
Gravitational Wave Astronomy sub-project C7.
We would like to thank Prof. Paul Kunitzsch (LMU Munich) for advise regarding the transliteration of the Arabic text.

{}

\end{document}